\documentclass[fleqn,usenatbib]{mnras}

\usepackage{newtxtext,newtxmath}

\usepackage[T1]{fontenc}

\DeclareRobustCommand{\VAN}[3]{#2}
\let\VANthebibliography\thebibliography
\def\thebibliography{\DeclareRobustCommand{\VAN}[3]{##3}\VANthebibliography}

\usepackage{graphicx}	
\usepackage{amsmath}	
\usepackage[utf8]{inputenc}

\newcommand{\software}[1]{\mbox{\sc {#1}}}

\title[EBLM~J0608-59]{Fundamental effective temperature measurements for eclipsing binary stars -- V. The circumbinary planet system EBLM J0608$-$59}

\author[P. F. L. Maxted et al.]{
P. F. L. Maxted$^{1\,\href{https://orcid.org/0000-0003-3794-1317}{\includegraphics[scale=0.5]{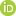}}}$
N. J. Miller,$^{2\,\href{https://orcid.org/0000-0001-9550-1198}{\includegraphics[scale=0.5]{orcid.jpg}}}$ 
D. Sebastian,$^{3\,\href{https://orcid.org/0000-0002-2214-9258}{\includegraphics[scale=0.5]{orcid.jpg}}}$ 
A. H. M. J. Triaud$^{3\,\href{https://orcid.org/0000-0002-5510-8751}{\includegraphics[scale=0.5]{orcid.jpg}}}$,
D.~V. Martin,$^{4,5\,\href{https://orcid.org/0000-0002-7595-6360}{\includegraphics[scale=0.5]{orcid.jpg}}}$
and 
A. Duck$^{5\,\href{https://orcid.org/0000-0002-4531-6899}{\includegraphics[scale=0.5]{orcid.jpg}}}$\\
$^{1}$Astrophysics Group, Keele University, Staffordshire ST5 5BG, UK\\
$^{2}$Centrum Astronomiczne im. Mikołaja Kopernika, Polish Academy of Sciences, Bartycka 18, 00-716, Warsaw, Poland\\
$^{3}$School of Physics and Astronomy, University of Birmingham, Edgbaston, Birmingham B15 2TT, UK\\
$^{4}$Department of Physics \& Astronomy, Tufts University, Medford, MA 02155, USA\\
$^{5}$Department of Astronomy, The Ohio State University, Columbus, OH 43210, USA\\
}
\date{Accepted XXX. Received YYY; in original form ZZZ}

\pubyear{2024}

\begin{document}
\label{firstpage}
\pagerange{\pageref{firstpage}--\pageref{lastpage}}
\maketitle

\begin{abstract}
EBLM~J0608$-$59 / TOI-1338 / BEBOP-1 is a 12$^{\rm th}$-magnitude, F9\,V star in an eclipsing binary with a much fainter M~dwarf companion on a wide, eccentric orbit (P=14.6 d). 
The binary is orbited by two circumbinary planets: one transiting on a 95-day orbit and one non-transiting on a 215-day orbit.
We have used high-precision photometry from the TESS mission combined with direct mass measurements for the two stars published recently to measure the following model-independent radii:  $R_1 = 1.32  \pm 0.02  R_{\odot}$, $R_2 = 0.309  \pm 0.004 R_{\odot}$.
Using $R_1$ and the parallax from Gaia EDR3 we find that this star's angular diameter is $\theta = 0.0309 \pm 0.0005$\,mas. 
The apparent bolometric flux of the primary star corrected for both extinction and the contribution from the M~dwarf ($<0.4$\,per~cent) is ${\mathcal F}_{\oplus,0} = (0.417\pm 0.005)\times10^{-9}$\,erg\,cm$^{-2}$\,s$^{-1}$. 
Hence, this F9\,V star has an effective temperature $T_{\rm eff,1} = 6031{\rm\,K} \pm 46{\rm \,K\,(rnd.)} \pm 10 {\rm \,K\,(sys.)}$. 
EBLM~J0608$-$59 is an ideal benchmark star that can be added to the sample of such systems we are establishing for ``end-to-end'' tests of the stellar parameters measured by large-scale spectroscopic surveys.  
\end{abstract}




\section{Introduction}

EBLM~J0608$-$59 / TOI-1338 / BEBOP-1 is an F9\,V star with an M-dwarf companion star that transits the primary star once every 14.6 days. 
The eclipses were found in ground-based photometry by the WASP survey \citep[Wide Angle Search for Planets,][]{2006PASP..118.1407P}. 
Radial velocity measurements showed that the companion is too massive to be an exoplanet and so it was flagged as an EBLM system, i.e. an eclipsing binary with a low-mass companion \citep{2023Univ....9..498M}. 
\cite{2017A&A...608A.129T} published a spectroscopic orbit for this SB1 (single-lined) binary system along with over 100 other EBLM systems identified by the WASP project. 
Additional high-precision radial velocity (RV) measurements were reported as part of the BEBOP survey \citep[Binaries Escorted By Orbiting Planets,][]{2019A&A...624A..68M}. 
These RV measurements combined with the RV measurements reported in \cite{2017A&A...608A.129T} were used to set an upper limit $M\approx 1M_{\rm Jup}$ to the mass of any circumbinary planets with periods of a few hundred days orbiting this binary system. 
Nevertheless, transits from a circumbinary planet with a radius $\approx 7M_{\oplus}$ were discovered in the light curve of this star (TOI-1338) from the Transiting Exoplanet Survey Satellite (TESS) mission by \citet{2020AJ....159..253K}, the first such system identified from TESS photometry. 
\citet{2023NatAs...7..702S} have reported the discovery of a second circumbinary planet, BEBOP-1\,c, with a minimum mass of  65\,M$_{\oplus}$ and an orbital  period of 216\,days based on additional RV measurements obtained with the HARPS and ESPRESSO spectrographs  \citep{,2002Msngr.110....9P, 2021A&A...645A..96P}. 
\citeauthor{2023NatAs...7..702S} found that the transiting circumbinary planet, BEBOP-1\,b, has a mass  $\loa 22\,M_{\oplus}$ based on their analysis of all the available high-precision RV data to-date.

Although the M-dwarf companion contributes less than 0.5\,per~cent of the flux at optical wavelengths, \citet{2024MNRAS.tmp.1016S} were able to detect the spectrum of this faint companion star by using over 100 spectra of EBLM~J0608$-$59 obtained with the ESPRESSO spectrograph on the 8.2-m VLT telescope. This detection combined with the spectroscopic orbit of the F9\,V primary star leads to a direct mass measurement for both stars, viz $M_1 = 1.098 \pm 0.017 M_{\odot}$ and  $M_2 =0.307 \pm 0.003\,M_{\odot}$. 

In this study, we combine the masses of the two stars measured by \cite{2024MNRAS.tmp.1016S} with an analysis of the TESS light curve to obtain precise and accurate measurements for the F9\,V primary and M-type secondary star in EBLM~J0608$-$59. We then use the parallax of the system from Gaia DR3 \citep{2016A&A...595A...1G,2023A&A...674A...1G} combined with flux measurements at optical and infrared wavelengths to measure directly the effective temperature (T$_{\rm eff}$) of the F9\,V star. This direct, accurate and precise measurement of T$_{\rm eff}$ for a moderately bright star ($V\approx12$) in an eclipsing binary system that has a negligible flux contribution from its M-dwarf companion makes EBLM~J0608$-$59 an ideal benchmark star that can be used for ``end-to-end'' tests of stellar parameters obtained from spectroscopic surveys, i.e. a direct comparison of the T$_{\rm eff}$ and $\log g$  value published by large spectroscopic surveys to accurate,  independent measurements of these quantities for a benchmark star observed in the same way as other stars in the survey.

\section{Analysis}

\subsection{TESS photometry}
EBLM~J0608$-$59 has been observed at 120-s cadence by TESS \citep{2015JATIS...1a4003R} in 30 sectors over 3 years. 
The light curves available from the TESS Science Processing Operations Centre (SPOC) show systematic errors in some sectors that appear to be due to the transit interfering with the de-trending of the light curves. 
We therefore decided to produce our own light curves from the target pixel files (TPFs) using the package \software{Lightkurve 2.0} \citep{2018ascl.soft12013L}.\footnote{\url{https://docs.lightkurve.org/}}
The TPFs were downloaded from the Mikulski Archive for Space Telescopes\footnote{\url{https://archive.stsci.edu/}} (MAST) and light curves computed using the same target and background apertures used for the generation of the SPOC light curves. 
The main source of systematic error in these light curves appears to be inaccurate correction for the background flux. 
We therefore decided to use \software{RegressionCorrector} function in \software{Lightkurve} to perform a linear regression sector-by-sector against the estimated background flux excluding transits and eclipses from the calculation of the model. 
This model was then evaluated at all times of observation in each sector and subtracted from the total flux in the target aperture.
We then extracted the sections of the light curve within one full eclipse width of either a primary or secondary eclipse and divided each of these sections by a straight line fit by least squares to the data either side of the eclipse. 
These sections of the light curve were grouped into 49 subsets containing 1 or 2 primary eclipses and 1 secondary eclipse. 

We performed a least-squares fit of the Nelson-Davis-Etzel (NDE) light curve model \citep{1972ApJ...174..617N}  independently to each of the 49 subsets of TESS photometry data using \software{jktebop}\footnote{\url{http://www.astro.keele.ac.uk/jkt/codes/jktebop.html}} \citep{2010MNRAS.408.1689S}. 
The stars are very nearly spherical and we have not used data between the eclipses so the ellipsoidal effect was ignored in the analysis of the light curves. 
The reflection effect was also ignored. 
Limb darkening was modelled using the power-2 law recently implemented in {\sc jktebop} \citep{2023Obs...143...71S}. 
The values of the parameters $h_1$ and $h_2$ were estimated by interpolation within the relevant table from \citet{2018A&A...616A..39M}.
The effect of the assumed value of $h_1$ for the primary star can be seen in the curvature of the light curve at the bottom of the primary eclipse so this parameter was allowed to vary in the analysis of the light curve. 
The effect of $h_2$ on the light curve model is very subtle so this parameter was fixed at the value obtained from  \citet{2018A&A...616A..39M}. 
The assumed limb darkening of the secondary star has a negligible effect on the light curve so we used the fixed nominal values $h_1=0.7$ and $h_2=0.4$.

The free parameters in the least-squares fit to the light curve are: the sum of the stellar radii in units of
the semi-major axis (fractional radii), $r_1+r_2=(R_1+R_2)/a$; the ratio of the stellar radii, $k=R_2/R_1$; the ratio of the surface brightness at the centre of each stellar disc, $J_0$; the orbital inclination, $i$; the time of mid-primary eclipse, $T_0$;  $e\sin(\omega)$ and $e\cos(\omega)$, where $e$ is the orbital eccentricity and $\omega$ is the longitude of periastron for the primary star; ``third light'', $\ell_3$. 
The orbital period was fixed at the value $P=14.6085577$. We also included priors on the values of $e\sin(\omega) = 0.137609\pm 0.000034$ and $e\cos(\omega)=-0.072464 \pm 0.000043$ based on the spectroscopic orbit of the primary star from \citet{2024MNRAS.tmp.1016S}. 
The value for the contamination of the photometric aperture provided in the meta data for the SPOC light curves is 1\,per~cent so we assume $\ell_3 = 0.01 \pm 0.01$ as a prior in the fit.

\begin{figure}
    \centering
    \includegraphics[width=1\linewidth]{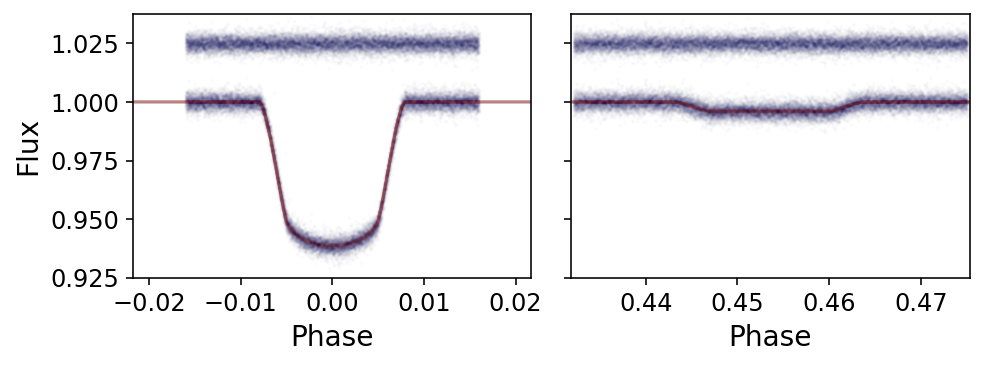}
    \caption{TESS photometry of EBLM~J0608$-$59 as a function of orbital phase (blue points) and best-fit light curve models for these data fitted as 49 individual subsets (red lines). The residuals from the best-fit models are shown offset vertically above the light curve data.}
    \label{fig:lcfit}
\end{figure}

\subsection{Eclipse ephemeris}
We combined the 49 times of mid-primary transit from our analysis of the TESS photometry 
with 4 times of mid-primary transit determined from a least-squares fit to data from the WASP project to update our estimate of the orbital period. The data from the WASP project were fit with the same model as the TESS data but with only $J_0$, $T_0$, $\ell_3$ as free parameters, the other parameters being fixed at values determined from a fit to the TESS light curve. All the data from one observing season were fit together to determine each of the 4 times of mid-primary eclipse. All the times of mid-transit used in this analysis are provided in the supplementary material available with the on-line version of this manuscript.

\begin{table*}
\centering
\caption{Observed apparent magnitudes for EBLM~J0608$-$59 and predicted values based on our synthetic photometry. The predicted magnitudes are shown with error estimates from the uncertainty on the zero-points for each photometric system. The pivot wavelength for each band pass is shown in the column headed $\lambda_{\rm pivot}$. The magnitudes of the primary G0V star alone corrected for the contribution to the total flux from the M-dwarf are shown in the column headed $m_1$. The flux ratio in each band is shown in the final column.}
\label{tab:mags}
\begin{tabular}{@{}lrrrrrr} 
\hline
		Band &  $\lambda_{\rm pivot}$ [nm]& \multicolumn{1}{c}{Observed} & \multicolumn{1}{c}{Computed} & 
\multicolumn{1}{c}{$\rm O-\rm C$} & \multicolumn{1}{c}{$m_1$} & \multicolumn{1}{c}{$\ell$} [\%]\\
\hline
BP    &  5110 &$ 12.155\pm 0.003$&$ 12.137\pm 0.003$&$ +0.018\pm 0.004$&$ 12.156 \pm  0.003 $&  0.02  \\
G     &  6218 &$ 11.873\pm 0.003$&$ 11.863\pm 0.003$&$ +0.009\pm 0.004$&$ 11.873 \pm  0.003 $&  0.09  \\
RP    &  7769 &$ 11.426\pm 0.004$&$ 11.435\pm 0.004$&$ -0.009\pm 0.005$&$ 11.428 \pm  0.004 $&  0.19  \\
u     &  3493 &$ 13.447\pm 0.012$&$ 13.536\pm 0.030$&$ -0.088\pm 0.032$&$ 13.447 \pm  0.012 $&  0.00  \\
v     &  3836 &$ 13.078\pm 0.012$&$ 13.075\pm 0.020$&$ +0.003\pm 0.023$&$ 13.078 \pm  0.012 $&  0.00  \\
J     & 12406 &$ 10.950\pm 0.023$&$ 10.950\pm 0.005$&$ -0.000\pm 0.024$&$ 10.958 \pm  0.023 $&  0.78  \\
H     & 16490 &$ 10.695\pm 0.022$&$ 10.679\pm 0.005$&$ +0.016\pm 0.023$&$ 10.707 \pm  0.022 $&  1.06  \\
K$_s$ & 21629 &$ 10.635\pm 0.021$&$ 10.604\pm 0.005$&$ +0.031\pm 0.022$&$ 10.649 \pm  0.021 $&  1.29  \\
W1    & 33683 &$ 10.570\pm 0.023$&$ 10.534\pm 0.002$&$ +0.036\pm 0.023$&$ 10.586 \pm  0.023 $&  1.49  \\
W2    & 46179 &$ 10.597\pm 0.020$&$ 10.522\pm 0.002$&$ +0.075\pm 0.020$&$ 10.617 \pm  0.020 $&  1.83  \\
W3    &120731 &$ 10.594\pm 0.043$&$ 10.632\pm 0.002$&$ -0.038\pm 0.043$&$ 10.621 \pm  0.044 $&  2.44  \\
\hline
\end{tabular}
\end{table*}

\begin{table}
	\centering
	\caption{Colour-$T_{\rm eff}$ relations used to establish Gaussian priors on the flux ratio at infrared wavelengths for EBLM~J0608$-$59. The dependent variables are $X_1 = T_{\rm eff,1}-6\,{\rm kK}$ and $X_2 = T_{\rm eff,2}-3.3\,{\rm kK}$. }
	\label{tab:frp}
	\begin{tabular}{@{}lll} 
		\hline
		Colour &  Primary & Secondary  \\
		\hline
V$-$J           & $ 1.089 -0.4370 \, X_1 \pm 0.015 $ & $ 4.187 -2.762 \, X_2 \pm  0.11 $ \\
V$-$H           & $ 1.343 -0.5838 \, X_1 \pm 0.019 $ & $ 4.776 -2.552 \, X_2 \pm  0.15 $ \\
V$-$K$_{\rm s}$ & $ 1.421 -0.6145 \, X_1 \pm 0.017 $ & $ 5.049 -2.776 \, X_2 \pm  0.12 $ \\
V$-$W1          & $ 1.469 -0.6185 \, X_1 \pm 0.027 $ & $ 5.207 -2.720 \, X_2 \pm  0.12 $ \\
V$-$W2          & $ 1.473 -0.6138 \, X_1 \pm 0.045 $ & $ 5.365 -2.957 \, X_2 \pm  0.11 $ \\
V$-$W3          & $ 1.421 -0.6284 \, X_1 \pm 0.023 $ & $ 5.477 -3.091 \, X_2 \pm  0.13 $ \\
\hline
\end{tabular}
\end{table}

\subsection{Mass, radius and luminosity}
The mass, radius of the two stars with their standard errors have been computed in nominal solar units \citep{2015arXiv151007674M} using a Monte Carlo simulation assuming independent Gaussian error on the values of $r_1+r_2$, $k=r_2/r_1$, $i$ and $e$ from the fit the TESS light curve and the values of $K_1$ and $K_2$ from \cite{2024MNRAS.tmp.1016S}. 
The error on $P=14.60855755$\,d is assumed to be negligible. 
The same Monte Carlo simulation is used to calculate the luminosity of the two stars and their standard errors taking random values of the effective temperatures, T$_{\rm eff,1}$ and T$_{\rm eff,2}$, sampled from the posterior probability distribution computed as described in the Section~\ref{sec:teb}.

\subsection{Effective temperature measurements}
\label{sec:teb}
The effective temperature for a star with Rosseland radius $R$ and total luminosity $L$ is defined by the equation  
\[L=4\pi R^2 \sigma_{\rm SB} {\rm T}_{\rm eff}^4,\]
where $\sigma_{\rm SB}$ is the Stefan-Boltzmann constant. For a binary star at distance $d$, i.e. with parallax $\varpi=1/d$, the flux corrected for extinction observed at the top of Earth's atmosphere is
\[f_{0,b}= f_{0,1}+f_{0,2}=\frac{\sigma_{\rm SB}}{4}\left[\theta_1^2{\rm T}_{\rm eff,1}^4 + \theta_2^2{\rm T}_{\rm eff,2}^4\right],\]
where $\theta_1=2R_1\varpi$ is the angular diameter of star 1, and similarly for star 2. All the quantities are known or can be measured for EBLM~J0608$-$59 provided we can accurately integrate the observed flux distributions for the two stars independently. 
This is possible because photometry of the combined flux from both stars is available from near-ultraviolet to mid-infrared wavelengths, and the flux ratio in the TESS band is measured accurately from our analysis of the light curve. 
Although we have no direct measurement of the flux ratio at infrared wavelengths, we can make a reasonable estimate for the small contribution of the M-dwarf to the measured total infrared flux using empirical colour\,--\,$T_{\rm eff}$ relations. 
The M-dwarf contributes less than 0.5\,per~cent to the total flux so it is not necessary to make a very accurate estimate of the M-dwarf flux distribution in order to derive an accurate value of $T_{\rm eff}$ for the F9\,V primary star.

The photometry used in this analysis is given in Table~\ref{tab:mags}. 
The Gaia photometry is from Gaia data release EDR3. J, H and Ks magnitudes are from the 2MASS survey \citep{2006AJ....131.1163S}. 
WISE magnitudes are from the All-Sky Release Catalog \citep{2012yCat.2311....0C} with corrections to Vega magnitudes made as recommended by \citet{2011ApJ...735..112J}.  
Details of the zero-points and response functions used to calculate synthetic photometry for these surveys from an assumed spectral energy distribution are given in \citet{2020MNRAS.497.2899M}. 
The u- and v-band photometry are taken from the SkyMapper survey DR4 \citep{2024arXiv240202015O}.

\begin{figure}
	\includegraphics[width=\columnwidth]{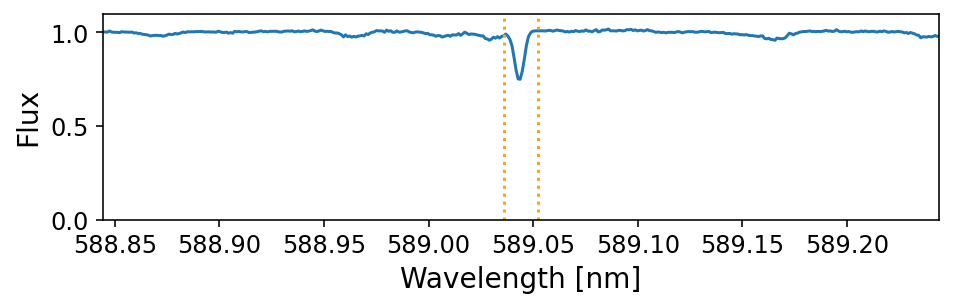}
	\includegraphics[width=\columnwidth]{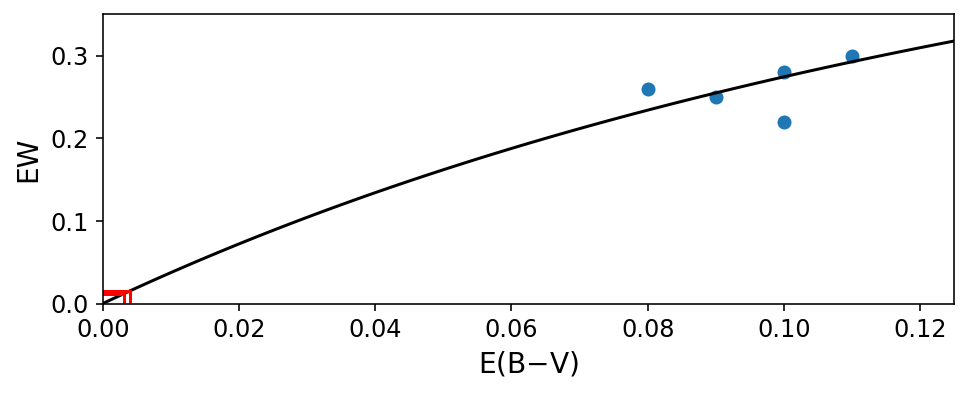}
    \caption{\label{fig:EW_NaI}
    Upper panel: The interstellar Na\,I~D$_1$ absorption feature in the ESPRESSO spectrum of EBLM~J0608$-$59. Vertical dashed lines shown the wavelength range used to measure EW(Na\,I~D$_1$). Lower panel: EW(Na\,I~D$_1$) -- E(B$-$V) relation from \protect{\citet{1997A&A...318..269M}} (solid line). The points are the calibrating data from \protect{\citeauthor{1997A&A...318..269M}} 
     with E(B$-$V) $<0.2$ that we use to assess the scatter in this relation at low reddening values. The red lines in the lower-left corner show the 1-$\sigma$ range of our  EW(Na\,I~D$_1$) for EBLM~J0608$-$59 and the inferred 1-$\sigma$ range of E(B$-$V).}
\end{figure}

To estimate the reddening towards EBLM~J0608$-$59 we use the calibration of E(B$-$V) versus the equivalent width of the interstellar Na\,I~D$_1$ line by \citet{1997A&A...318..269M}. 
This is an semi-empirical relation based on an analytic relation between equivalent width and column density assuming a constant gas-to-dust ratio with the constant of proportionality calibrated using using 18 stars that show a single interstellar Na\,I absorption line.
To measure EW(Na\,I~D$_1$) we selected 69 of the spectra obtain with the ESPRESSO spectrograph observed through an airmass $<1.4$. 
We first shifted these spectra into the rest frame of the primary star and then took the median value at each wavelength to obtain a high signal-to-noise spectrum of the F9\,V primary star. 
We then divided each observed spectrum by this spectrum of the F9\,V primary star after shifting it back to the barycentric rest frame. 
We then took the median of these residual spectra to obtain the high signal-to-noise spectrum of the interstellar features  shown in Fig.~\ref{fig:EW_NaI}. 
The equivalent width of the Na\,I~D$_1$ line measured by numerical integration is EW(Na\,I~D$_1) = 14 \pm 2$\,m\AA. 
This value is less than the values of EW(Na\,I~D$_1$) for all the stars in the  calibration sample of \citet{1997A&A...318..269M}. 
To estimate the uncertainty on this value of E(B$-$V) we take the sample standard deviation for the 5 stars in the calibration sample with the lowest values of  EW(Na\,I~D$_1)\approx 250$\,m\AA (Fig.~\ref{fig:EW_NaI}).
As can be seen in Fig.~\ref{fig:EW_NaI}, this is quite pessimistic given the very low reddening we infer for EBLM~J0608$-$59.
Based on this analysis we obtain the estimate E(B$-$V$)= 0.0036 \pm 0.0018$. 
We use this as a Gaussian prior in our analysis but exclude negative values of E(B$-$V).  
The co-added ESPRESSO spectra can be obtained from the supplementary online information that accompanies this article or from the corresponding author's website.\footnote{\url{https://www.astro.keele.ac.uk/pflm/BenchmarkDEBS/}} 

To establish colour\,--\,$T_{\rm eff}$ relations suitable for dwarf stars with $3100\,{\rm K} < {\rm T}_{\rm eff} < 3500\,{\rm K} $ we use a robust linear fit to the stars listed in Table~6  of \citet{2018MNRAS.475.1960F} within this $T_{\rm eff}$ range. 
Photometry for these stars is taken from the TESS input catalogue. 
To estimate a suitable standard error for a Gaussian prior based on this fit we use 1.25$\times$ the mean absolute deviation of the residuals from the fit. 
Colour\,--\,$T_{\rm eff}$ relations suitable for the primary F9\,V star were calculated in a similar way based on stars selected from the Geneva-Copenhagen survey \citep{2009A&A...501..941H, 2011A&A...530A.138C}  with $5750\,{\rm K} < {\rm T}_{\rm eff} < 6250\,{\rm K}$, $E({\rm B}-{\rm V})<0.05$ and $3.5 < \log g < 4.5$. 
The results are given in Table~\ref{tab:frp}.

The method we have developed to measure $T_{\rm eff}$ for eclipsing binary stars is described fully in \citet{2020MNRAS.497.2899M}.
Briefly, we use  \software{emcee} \citep{2013PASP..125..306F} to sample the posterior probability distribution $P(\Theta| D)\propto P(D|\Theta)P(\Theta)$ for the model parameters $\Theta$ with prior $P(\Theta)$ given the data, $D$ (observed apparent magnitudes and flux ratios). 
The model parameters are  $$\Theta = \left({\rm T}_{\rm eff,1},  {\rm T}_{\rm eff,2}, \theta_1, \theta_2, {\rm E}({\rm B}-{\rm V}), \sigma_{\rm ext}, \sigma_{\ell},  c_{1,1}, \dots, c_{2,1}, \dots\right).$$  
The prior $P(\Theta)$ is calculated using the angular diameters $\theta_1$ and $\theta_2$ derived from the radii $R_1$ and $R_2$ and the parallax $\varpi$, the priors on the flux ratio at infrared wavelengths based on the colour\,--\,T${\rm eff}$ relations in Table~\ref{tab:frp}, and the Gaussian prior on the reddening described above. 
The hyper-parameters $\sigma_{\rm ext}$ and $\sigma_{\ell}$ account for additional uncertainties in the synthetic magnitudes  and flux ratio, respectively, due to errors in zero-points, inaccurate response functions, stellar variability, etc. 
The parallax is taken from Gaia EDR3 with corrections to the zero-point from \citet{2022MNRAS.509.4276F}.

To calculate the synthetic photometry for a given value of $T_{\rm eff}$ we used a model spectral energy distribution (SED) multiplied by a distortion function, $\Delta(\lambda)$. The distortion function is a linear superposition of Legendre polynomials in log wavelength. The coefficients of the distortion function for star 1 are $c_{1,1}, c_{1,2}, \dots$, and similarly for star 2. The distorted SED for each star is normalized so that the total apparent flux prior to applying reddening is $\sigma_{\rm SB}\theta^2{\rm T}_{\rm eff}^4/4$. These distorted SEDs provide a convenient function that we can integrate to calculate synthetic photometry that has realistic stellar absorption features, and where the overall shape can be adjusted to match the observed magnitudes from ultraviolet to infrared wavelengths. This means that the effective temperatures we derive are based on the integrated stellar flux and the star's angular diameter, not SED fitting.

For this analysis we use model SEDs computed from BT-Settl model atmospheres \citep{2013MSAIS..24..128A} obtained from the Spanish Virtual Observatory.\footnote{\url{http://svo2.cab.inta-csic.es/theory/newov2/index.php?models=bt-settl}} 
For the F9\,V star appropriate we use the model with  $T_{\rm eff,1}=6000$\,K, $\log g_1 = 4.0$, $[{\rm Fe/H}] = 0.0$ and   $[{\rm \alpha/Fe}] = 0.0$. 
For the reference SED for the M dwarf companion we assume $T_{\rm eff,1}=3000$\,K, $\log g_1 = 5.0$, and the same composition.
We experimented with distortion functions with 3, 4 and 5 coefficients per star and found the results to be very similar in all cases. 
The results presented here use three distortion coefficient per star because there is no improvement in the quality of the fit if we use a larger number of coefficients. 
The predicted apparent magnitudes including their uncertainties from errors in the zero-points for each photometric system are compared to the observed apparent magnitudes in Table~\ref{tab:mags}.

\begin{figure}
	\includegraphics[width=\columnwidth]{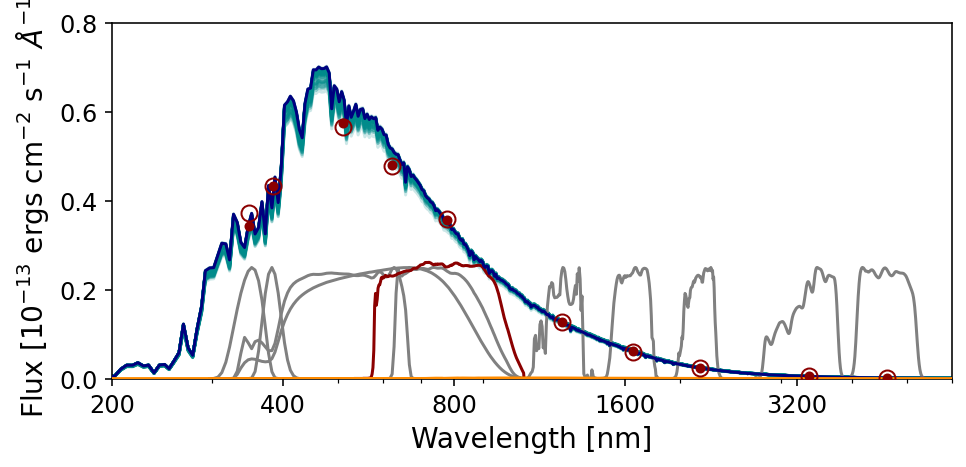}
    \caption{The spectral energy distribution (SED) of EBLM~J0608$-$59. The observed fluxes are plotted with open circles and the predicted fluxes for the mean of the posterior probability distribution (PPD) integrated over the response functions shown in grey are plotted with filled symbols. The SED predicted by the mean of the PPD is plotted in dark blue and light blue shows the SEDs produced from 100 random samples from the PPD. The contribution to the total SED from the M dwarf (barely visible) is shown in orange. The TESS response function is shown in red. The W3  mid-infrared bands also used in the analysis is not shown here. }
    \label{fig:sed}
\end{figure}

\section{Results}

The TESS light curve data and best-fit models for each subset are shown in Fig.~\ref{fig:lcfit}. The mean and standard error of the mean across the 49 subsets for each free parameter and some useful derived quantities are given in Table~\ref{tab:lcfit}.
The results from our least-squares fit of a linear ephemeris to the 49 times of mid-primary eclipse from the TESS data and 4  times of mid-primary eclipse from the WASP data are shown in the same table.

The posterior probability distribution for the model parameters from our analysis to measure the stellar effective temperatures is summarised in Table~\ref{tab:teb} and the spectral energy distribution is plotted in Fig.~\ref{fig:sed}.
The random errors quoted in Table~\ref{tab:teb} do not allow for the systematic error due to the uncertainty in the absolute calibration of the CALSPEC flux scale \citep{2014PASP..126..711B}. This additional systematic error is 10\,K for the G0V primary star and 7\,K for the M-dwarf companion. 

The mass, radius and luminosity of the two stars derived from these results are given in Table~\ref{tab:mr}. 
The two stars compared to other stars in eclipsing binary systems with accurate mass and radius measurements are shown in the Hertzsprung-Russell diagram in Fig.~\ref{fig:hrd}.

\begin{table}
\caption[]{Mean and standard error of the mean for free parameters and derived quantities from least-squares fits to 49 subsets of the TESS light curve of EBLM~J0608$-$59. }
\label{tab:lcfit}
\begin{center}
  \begin{tabular}{@{}lrl}
\hline
\noalign{\smallskip}
 \multicolumn{1}{@{}l}{Parameter} &
 \multicolumn{1}{l}{Value} &
 Note \\
\noalign{\smallskip}
\hline
\noalign{\smallskip}
BJD$_{\rm TDB}$ T$_0$ & $2459344.812081 \pm 0.000032 $ & Time of mid-transit \\
$P$    & $ 14.60855755 \pm 0.00000075 $ & days \\
$ r_1+r_2      $&$      0.0579 \pm      0.0007 $& \\
$ k=r_2/r_1    $&$     0.2336 \pm       0.0015 $& \\
$ h_1          $&$      0.804 \pm        0.015 $& \\
$ i            $&$     89\fdg6 \pm      0\fdg4 $& \\
$ \ell_3       $&$     0.0098 \pm      0.0007 $& Constrained by a prior.\\
$ e\cos(\omega)$&$   -0.072462 \pm    0.000005 $& Constrained by a prior. \\
$ e\sin(\omega)$&$  0.13760898 \pm  0.00000009 $& Constrained by a prior. \\
$ \ell_T       $&$      0.0041 \pm      0.0003 $& Flux ratio in TESS band\\
$ e            $&$    0.155522 \pm    0.000002 $& \\
$ J            $&$       0.075 \pm       0.006 $& $=\ell_T/k^2$\\
$ r_1          $&$      0.0469 \pm      0.0006 $& \\
$ r_2          $&$      0.01095 \pm      0.00014 $& \\
\noalign{\smallskip}
\hline
\end{tabular}
\end{center}
\end{table}

\begin{table}
\centering
\caption{Results from our analysis to obtain the effective temperatures for both stars in EBLM~J0608$-$59. }
\label{tab:teb}
\begin{tabular}{@{}lrrl} 
\hline
Parameter & 
\multicolumn{1}{l}{Value} & 
\multicolumn{1}{l}{Error} & Units \\
\hline
\noalign{\smallskip}
$T_{\rm eff,1}$ & $6031 $&$ \pm  46$ (rnd.) $\pm 10$ (sys.) & K \\
$T_{\rm eff,2}$ & $3220 $&$ \pm 135$ (rnd.) $\pm ~7$ (sys.) & K \\
$\theta_1$      & $ 0.0309 $&$ \pm 0.0005 $ & mas \\
$\theta_2$      & $ 0.0072 $&$ \pm 0.0001 $ & mas \\
E(B$-$V)        & $ 0.004 $&$ \pm 0.002 $ \\
$\sigma_{\rm ext}$  & $0.019 $&$ \pm 0.012 $ \\
$\sigma_{\ell} $ & $ 0.0017 $&$ \pm 0.0018 $ \\
$c_{1,1}$ & $  -0.271 $&$ \pm 0.088 $ \\
$c_{1,2}$ & $   0.174 $&$ \pm 0.058 $ \\
$c_{1,3}$ & $  -0.204 $&$ \pm 0.090 $ \\
$c_{2,1}$ & $   -0.14 $&$ \pm 0.45 $ \\
$c_{2,2}$ & $    0.08 $&$ \pm 0.37 $ \\
$c_{2,3}$ & $   -0.12 $&$ \pm 0.34 $ \\
\hline
\end{tabular}
\end{table}

\begin{table}
\caption[]{Fundamental parameters of the stars in EBLM~J0608$-$59. The metallicity [Fe/H] is taken from  \protect{\cite{2020AJ....159..253K}.}}

\label{tab:mr}
\begin{center}
  \begin{tabular}{lrrr}
\hline
\noalign{\smallskip}
 \multicolumn{1}{l}{Parameter} &
 \multicolumn{1}{l}{Value} &
 \multicolumn{1}{l}{Error} &
 \multicolumn{1}{r}{} \\
\noalign{\smallskip}
\hline
\noalign{\smallskip}
$M_1/{\mathcal M^{\rm N}_{\odot}}$&1.098  & $\pm$ 0.018 & [1.6 \%] \\
\noalign{\smallskip}
$M_2/{\mathcal M^{\rm N}_{\odot}}$&0.308 & $\pm$ 0.003 &[1.0 \%] \\
\noalign{\smallskip}
$M_2/M_1$ & $ 0.2800 $ &$ \pm 0.0017$ &  [0.62 \%] \\
\noalign{\smallskip}
$R_1/{\mathcal R^{\rm N}_{\odot}}$&1.321 & $\pm$ 0.017 &[1.5 \%] \\
\noalign{\smallskip}
$R_2/{\mathcal R^{\rm N}_{\odot}}$&0.309 & $\pm$ 0.004 &[1.7 \%] \\
\noalign{\smallskip}
$\log(T_{\rm eff,1}/{\rm K}) $  & $ 3.780 $ & $\pm$ 0.004  & [0.9 \%] \\
\noalign{\smallskip}
$\log(T_{\rm eff,2}/{\rm K)}$  & $ 3.506 $ & $\pm$ 0.018  & [4.0 \%] \\
\noalign{\smallskip}
$\rho_1/{\rho^{\rm N}_{\odot}}$  &0.477 & $\pm$ 0.017 &[3.7 \%] \\
\noalign{\smallskip}
$\rho_2/{\rho^{\rm N}_{\odot}}$  & 10.4 & $\pm$ 0.4 &[4.0 \%] \\
\noalign{\smallskip}
$\log g_1$ [cgs] & 4.24 & $\pm$ 0.01 & [2.5 \%]  \\
\noalign{\smallskip}
$\log g_2$ [cgs] & 4.95 & $\pm$ 0.01 & [2.6 \%] \\
\noalign{\smallskip}
$\log L_1/{\mathcal L^{\rm N}_{\odot}}$   & 0.32 & $\pm$ 0.02 & [4.3 \%]  \\
\noalign{\smallskip}
$\log L_2/{\mathcal L^{\rm N}_{\odot}}$   & $ -2.04 $& $\pm$ 0.07 & [16 \%]  \\
\noalign{\smallskip}
[Fe/H] & $0.01$ & $\pm$ 0.05  & \\ 
\noalign{\smallskip}
\hline
\end{tabular}
\end{center}
\end{table}

\section{Discussion}

\begin{figure}
    \centering
    \includegraphics[width=\linewidth]{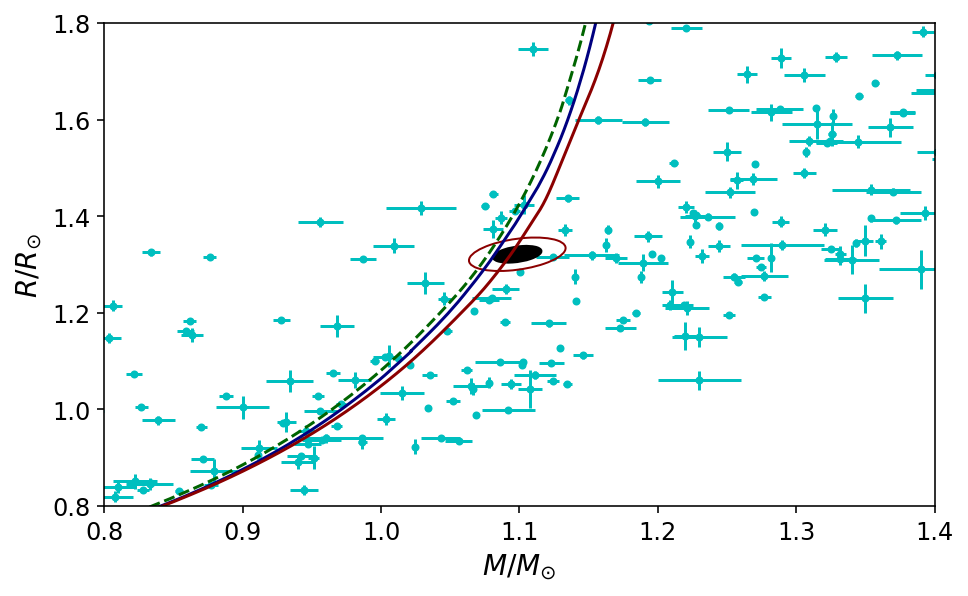}
    \includegraphics[width=\linewidth]{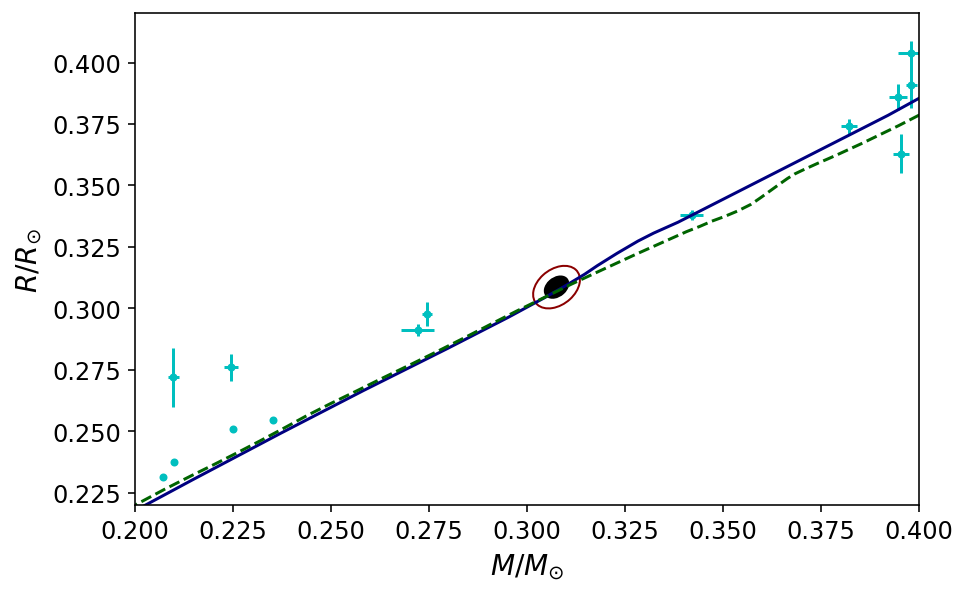}
    \includegraphics[width=\linewidth]{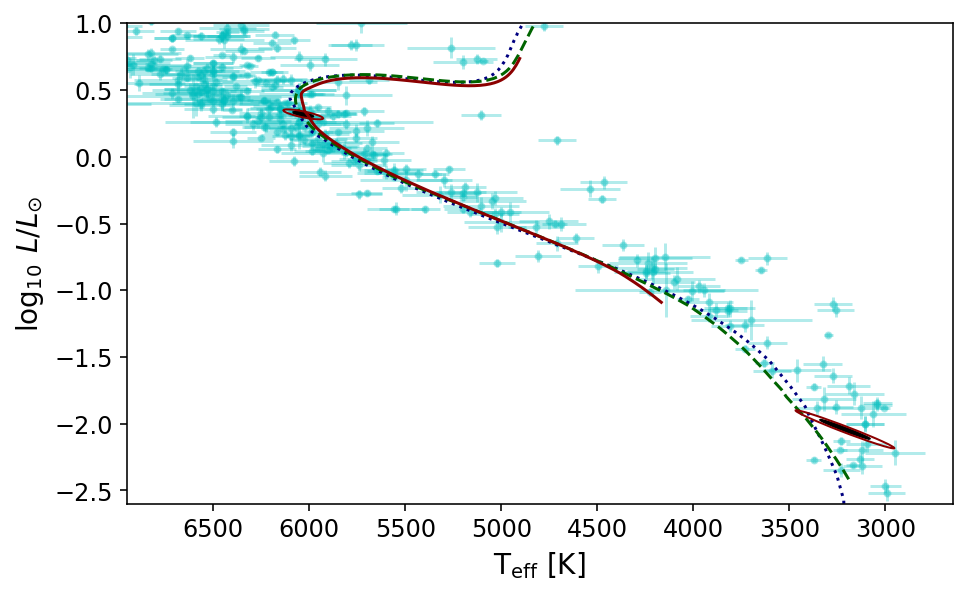}
    \caption{ Upper panel: primary component of EBLM~J0608$-$59 in the mass\,--\,radius plane. 
    Middle panel: secondary component of EBLM~J0608$-$59 in the mass\,--\,radius plane. 
    Lower panel: both components of EBLM~J0608$-$59 in the  Hertzsprung-Russell diagram.  
    The ellipses show 1-$\sigma$ and  2-$\sigma$ confidence regions on the parameters of EBLM~J0608$-$59. 
    All panels show isochrones for an age of 5.8\,Gyr assuming $[{\rm Fe/H}]=0.0$ from \software{bagemass} (red solid line), DSEP (blue dotted line) and MIST (green dashed line). 
    Cyan error bars show stars in eclipsing binary systems taken from DEBCat \citep{2015ASPC..496..164S}.}
    \label{fig:hrd}
\end{figure}

We used the software package \software{bagemass} version 1.3 \citep{2015A&A...575A..36M} to compare the parameters of the primary star, EBLM~J0608$-59$\,A, to a grid of stellar models computed with the  {\sc garstec} stellar evolution code \citep{2008Ap&SS.316...99W}. 
The methods used to calculate the stellar model grid are described in \citet{2013MNRAS.429.3645S}.  
A Markov-chain Monte-Carlo method to explore the posterior probability distribution (PPD) for the mass and age of a star based on its observed $T_{\rm eff}$, luminosity, mean stellar density and surface metal abundance [Fe/H]. 
Version 1.3 uses a Gaussian prior on the stellar mass rather than a power-law prior so that we can ensure that the mass derived is consistent with the observed value. 
We find a good fit to the observed properties of EBLM~J0608$-59$\,A for models with an age of $6.0 \pm 0.3$\,Gyr assuming a mixing length parameter $\alpha_{\rm MLT}=1.78$ (solar value) and solar helium abundance or slightly enhanced helium abundance (+0.02 dex). 
The quality of the fit for a reduced mixing length ($\alpha_{\rm MLT}=1.5$) is significantly worse. 
The best-fit isochrone with an age of 5.8\,Gyr assuming solar helium abundance and mixing length is shown in Fig.~\ref{fig:hrd}.
Isochrones for the same age and initial metal abundance from the Dartmouth stellar evolution database \citep[DSEP, ][]{2008ApJS..178...89D} and the MESA Isochrones \& Stellar Tracks \citep[MIST, ][]{2016ApJ...823..102C} are also shown in Fig.~\ref{fig:hrd}.

EBLM~J0608$-$59\,B matches very well the mass-radius relation for low-mass stars from the MIST and DSEP stellar model grids (Fig.~\ref{fig:hrd}). This is unusual for stars in this mass range, which tend to be larger than predicted by stellar models. These stars also tend to be cooler than predicted, which is also the case for EBLM~J0608$-$59\,B. These measurements are a valuable addition to the small sample of stars with independent mass and radius measurements around the mass range where there are  discontinuities in the relationships between the observed properties of M-dwarfs \citep{2018ApJ...861L..11J, 2019MNRAS.484.2674R}. These discontinuities are thought to be related to the transition between stars with a radiative core and fully-convective stars, but may also be associated with mixing of $^3$He during merger of envelope and core convection zones that occurs for masses $\approx 0.34\,M_{\odot}$ \citep{2018MNRAS.480.1711M}. It may then be worthwhile to improve on the precision of the T$_{\rm eff}$ measurement presented here using observations of the eclipse depth at near-infrared wavelengths, i.e. around the peak of the M-dwarf's SED.  

\cite{2020AJ....159..253K} derived the following masses and radii for the stars in EBLM~J$0608-59$: $M_1 = 1.127 \smash{^{+0.068}_{-0.069}}$,  $R_1 = 1.331 \smash{^{+0.024}_{-0.026}}\,R_{\odot}$, $M_2 = 0.313 \smash{^{+0.011}_{-0.012}}$ $R_2 = 0.3089\smash{^{+0.0056}_{-0.0060}} \,R_{\odot}$. The value of K$_2$ that we have used was not available at the time of that study so constraints on the primary star's radius and mass derived from the analysis of its spectral energy distribution and spectrum were required to obtain these mass and radius estimates.  The agreement between the values from this study and our results is excellent. The better precision in the values we have derived is partly because we have a precise measurement of the stellar masses from \cite{2024MNRAS.tmp.1016S}, and partly because we have been able to use light curve data from 3 cycles of observations with the TESS spacecraft whereas only 1 cycle of observations was available to \citeauthor{2020AJ....159..253K}. This gives us some reassurance that the parameters of M-dwarfs and other bodies orbiting Sun-like stars (e.g. brown dwarfs and planets) can be measured reliably using techniques such as those employed by \citeauthor{2020AJ....159..253K}.

We have investigated the impact of using the reddening estimate given in the catalogue published by \citet{2022A&A...658A..91A} instead of the prior on E(B$-$V) derived from the interstellar Na\,I~D$_1$ absorption line. 
This reddening estimate was computed with the StarHorse code \citep{2018MNRAS.476.2556Q} using Gaia EDR3 spectrophotometry and broad-band photometry from various optical and infrared surveys. If we use the prior $\rm E(\rm B-\rm V) = 0.058 \pm 0.036$ based on the value of A$_{\rm V}$ given by \citeauthor{2022A&A...658A..91A}, we obtain T$_{\rm eff,1} = 6214 \pm 115$\,K, T$_{\rm eff,2} =  3184 \pm 142$\,K. This is consistent with the effective temperatures given in  Table~\ref{tab:teb} within the rather broad errors that result from the larger uncertainty in this reddening estimate. 
The value of T$_{\rm eff} = 6272 \pm 173$\,K given in the same catalogue is consistent with the value of T$_{\rm eff,1}$ obtained using the reddening estimate from the same source. However, the value of E(B$-$V) from \citeauthor{2022A&A...658A..91A} corresponds to a value of  EW(Na\,I~D$_1) = 180  \smash{^{+67}_{-128}}$\,m\AA, which is an order-of-magnitude larger than the value we observe for this star. 
It would certainly be worthwhile to obtain suitable data to calibrate the E(B$-$V) -- EW(Na\,I~D$_1)$ relation shown in Fig.~\ref{fig:EW_NaI} for stars with little reddening, but this current calibration is clearly preferable to the weaker constraint on E(B$-$V) that can currently be obtained from spectrophotometry and broad-band photometry for nearby stars.

\section{Conclusion}
EBLM~J0608$-$59 adds to a small but growing sample of eclipsing binaries for which we have directly measured the effective temperature of a solar-type star with a much fainter M-dwarf companion \citep{2022MNRAS.513.6042M, 2023MNRAS.522.2683M}. 
These stars are complementary to benchmark stars for which the effective temperature estimate is based on the angular diameters ($\theta$) measured using interferometry. 
Interferometric measurements of $\theta$ are technically challenging and so suffer from systematic errors that can be seen in the poor agreement between measurements of the same star made with different instruments. 
For solar-type dwarf stars that typically have angular diameters for $\loa 1$\,mas, these systematic can be $\approx 10$\,per~cent or more \citep{2022A&A...658A..47K, 2024A&A...682A.145S}. 
The effective temperature measurements we have made from eclipsing binary stars have the advantage that the sources of systematic error are small and well-understood, and the surface gravity of the stars is known to very high accuracy. 
In addition, these stars are within the magnitude range (V$\sim 10$) that can be observed directly by instruments used for large-scale spectroscopic surveys in their standard observing modes. 
This makes it feasible to conduct ``end-to-end'' tests of the accuracy of stellar parameters derived by these surveys.
This is often not the case for the very bright stars that are currently accessible to interferometric measurements.

\section*{Acknowledgements}

We thank the anonymous referee for their comments that have improved the manuscript, particularly the suggestion to investigate the impact of using a different reddening estimate.

This research is supported work funded from the European Research
Council (ERC) under the European Union’s Horizon 2020 research
and innovation programme (grant agreement No. 803193 - BEBOP).

This research is supported work funded from the European Research Council (ERC) under the European Union’s Horizon 2020 research and innovation program (grant agreement No. 951549 - UniverScale);  the Polish National Science Center grants MAESTRO 2017/26/A/ST9/00446 and BEETHOVEN 2018/31/G/ST9/03050; the Polish Ministry of Science and Higher Education grant DIR/WK/2018/0.

This work has made use of data from the European Space Agency (ESA) mission
{\it Gaia} (\url{https://www.cosmos.esa.int/gaia}), processed by the {\it Gaia}
Data Processing and Analysis Consortium (DPAC,
\url{https://www.cosmos.esa.int/web/gaia/dpac/consortium}). Funding for the DPAC
has been provided by national institutions, in particular the institutions
participating in the {\it Gaia} Multilateral Agreement.

This paper includes data obtained through the TESS Guest investigator programs  G06022 (PI Martin), G05024 (PI Martin), G04157 (PI Martin), G03216 (PI Martin) and G022253 (PI Martin).

This paper includes data collected by the TESS mission, which is publicly available from the Mikulski Archive for Space Telescopes (MAST) at the Space Telescope Science Institure (STScI). Funding for the TESS mission is provided by the NASA Explorer Program directorate. STScI is operated by the Association of Universities for Research in Astronomy, Inc., under NASA contract NAS 5–26555. We acknowledge the use of public TESS Alert data from pipelines at the TESS Science Office and at the TESS Science Processing Operations Center.

This research made use of Lightkurve, a Python package for Kepler and TESS data analysis \citep{2018ascl.soft12013L}.

Based on data obtained from the ESO Science Archive Facility with DOI(s) :  https://doi.eso.org/10.18727/archive/21. 
\section*{Data Availability}
The data underlying this article are available in the following repositories:  
Mikulski Archive for Space Telescopes -- \url{https://archive.stsci.edu/}  (TESS); ESO Science Archive Facility -- \url{https://archive.eso.org/}.



\bibliographystyle{mnras}
\bibliography{allbib} 

\begin{thebibliography}{}
\makeatletter
\relax
\def\mn@urlcharsother{\let\do\@makeother \do\$\do\&\do\#\do\^\do\_\do\%\do\~}
\def\mn@doi{\begingroup\mn@urlcharsother \@ifnextchar [ {\mn@doi@}
  {\mn@doi@[]}}
\def\mn@doi@[#1]#2{\def\@tempa{#1}\ifx\@tempa\@empty \href
  {http://dx.doi.org/#2} {doi:#2}\else \href {http://dx.doi.org/#2} {#1}\fi
  \endgroup}
\def\mn@eprint#1#2{\mn@eprint@#1:#2::\@nil}
\def\mn@eprint@arXiv#1{\href {http://arxiv.org/abs/#1} {{\tt arXiv:#1}}}
\def\mn@eprint@dblp#1{\href {http://dblp.uni-trier.de/rec/bibtex/#1.xml}
  {dblp:#1}}
\def\mn@eprint@#1:#2:#3:#4\@nil{\def\@tempa {#1}\def\@tempb {#2}\def\@tempc
  {#3}\ifx \@tempc \@empty \let \@tempc \@tempb \let \@tempb \@tempa \fi \ifx
  \@tempb \@empty \def\@tempb {arXiv}\fi \@ifundefined
  {mn@eprint@\@tempb}{\@tempb:\@tempc}{\expandafter \expandafter \csname
  mn@eprint@\@tempb\endcsname \expandafter{\@tempc}}}

\bibitem[\protect\citeauthoryear{{Allard}, {Homeier}, {Freytag},
  {Schaffenberger}  \& {Rajpurohit}}{{Allard}
  et~al.}{2013}]{2013MSAIS..24..128A}
{Allard} F.,  {Homeier} D.,  {Freytag} B.,  {Schaffenberger} W.,   {Rajpurohit}
  A.~S.,  2013, Memorie della Societa Astronomica Italiana Supplementi, \href
  {https://ui.adsabs.harvard.edu/abs/2013MSAIS..24..128A} {24, 128}

\bibitem[\protect\citeauthoryear{{Anders} et~al.,}{{Anders}
  et~al.}{2022}]{2022A&A...658A..91A}
{Anders} F.,  et~al., 2022, \mn@doi [\aap] {10.1051/0004-6361/202142369}, \href
  {https://ui.adsabs.harvard.edu/abs/2022A&A...658A..91A} {658, A91}

\bibitem[\protect\citeauthoryear{{Bohlin}, {Gordon}  \& {Tremblay}}{{Bohlin}
  et~al.}{2014}]{2014PASP..126..711B}
{Bohlin} R.~C.,  {Gordon} K.~D.,   {Tremblay} P.~E.,  2014, \mn@doi [\pasp]
  {10.1086/677655}, \href
  {https://ui.adsabs.harvard.edu/abs/2014PASP..126..711B} {126, 711}

\bibitem[\protect\citeauthoryear{{Casagrande}, {Sch{\"o}nrich}, {Asplund},
  {Cassisi}, {Ram{\'\i}rez}, {Mel{\'e}ndez}, {Bensby}  \&
  {Feltzing}}{{Casagrande} et~al.}{2011}]{2011A&A...530A.138C}
{Casagrande} L.,  {Sch{\"o}nrich} R.,  {Asplund} M.,  {Cassisi} S.,
  {Ram{\'\i}rez} I.,  {Mel{\'e}ndez} J.,  {Bensby} T.,   {Feltzing} S.,  2011,
  \mn@doi [\aap] {10.1051/0004-6361/201016276}, \href
  {https://ui.adsabs.harvard.edu/abs/2011A&A...530A.138C} {530, A138}

\bibitem[\protect\citeauthoryear{{Choi}, {Dotter}, {Conroy}, {Cantiello},
  {Paxton}  \& {Johnson}}{{Choi} et~al.}{2016}]{2016ApJ...823..102C}
{Choi} J.,  {Dotter} A.,  {Conroy} C.,  {Cantiello} M.,  {Paxton} B.,
  {Johnson} B.~D.,  2016, \mn@doi [\apj] {10.3847/0004-637X/823/2/102}, \href
  {https://ui.adsabs.harvard.edu/abs/2016ApJ...823..102C} {823, 102}

\bibitem[\protect\citeauthoryear{{Cutri} \& {et al.}}{{Cutri} \& {et
  al.}}{2012}]{2012yCat.2311....0C}
{Cutri} R.~M.,  {et al.} 2012, VizieR Online Data Catalog, \href
  {https://ui.adsabs.harvard.edu/abs/2012yCat.2311....0C} {p. II/311}

\bibitem[\protect\citeauthoryear{{Dotter}, {Chaboyer}, {Jevremovi{\'c}},
  {Kostov}, {Baron}  \& {Ferguson}}{{Dotter}
  et~al.}{2008}]{2008ApJS..178...89D}
{Dotter} A.,  {Chaboyer} B.,  {Jevremovi{\'c}} D.,  {Kostov} V.,  {Baron} E.,
  {Ferguson} J.~W.,  2008, \mn@doi [\apjs] {10.1086/589654}, \href
  {https://ui.adsabs.harvard.edu/abs/2008ApJS..178...89D} {178, 89}

\bibitem[\protect\citeauthoryear{{Flynn}, {Sekhri}, {Venville}, {Dixon},
  {Duffy}, {Mould}  \& {Taylor}}{{Flynn} et~al.}{2022}]{2022MNRAS.509.4276F}
{Flynn} C.,  {Sekhri} R.,  {Venville} T.,  {Dixon} M.,  {Duffy} A.,  {Mould}
  J.,   {Taylor} E.~N.,  2022, \mn@doi [\mnras] {10.1093/mnras/stab3156}, \href
  {https://ui.adsabs.harvard.edu/abs/2022MNRAS.509.4276F} {509, 4276}

\bibitem[\protect\citeauthoryear{{Foreman-Mackey}, {Hogg}, {Lang}  \&
  {Goodman}}{{Foreman-Mackey} et~al.}{2013}]{2013PASP..125..306F}
{Foreman-Mackey} D.,  {Hogg} D.~W.,  {Lang} D.,   {Goodman} J.,  2013, \mn@doi
  [\pasp] {10.1086/670067}, \href
  {https://ui.adsabs.harvard.edu/abs/2013PASP..125..306F} {125, 306}

\bibitem[\protect\citeauthoryear{{Fouqu{\'e}} et~al.,}{{Fouqu{\'e}}
  et~al.}{2018}]{2018MNRAS.475.1960F}
{Fouqu{\'e}} P.,  et~al., 2018, \mn@doi [\mnras] {10.1093/mnras/stx3246}, \href
  {https://ui.adsabs.harvard.edu/abs/2018MNRAS.475.1960F} {475, 1960}

\bibitem[\protect\citeauthoryear{{Gaia Collaboration} et~al.,}{{Gaia
  Collaboration} et~al.}{2016}]{2016A&A...595A...1G}
{Gaia Collaboration} et~al., 2016, \mn@doi [\aap]
  {10.1051/0004-6361/201629272}, \href
  {https://ui.adsabs.harvard.edu/abs/2016A&A...595A...1G} {595, A1}

\bibitem[\protect\citeauthoryear{{Gaia Collaboration} et~al.,}{{Gaia
  Collaboration} et~al.}{2023}]{2023A&A...674A...1G}
{Gaia Collaboration} et~al., 2023, \mn@doi [\aap]
  {10.1051/0004-6361/202243940}, \href
  {https://ui.adsabs.harvard.edu/abs/2023A&A...674A...1G} {674, A1}

\bibitem[\protect\citeauthoryear{{Holmberg}, {Nordstr{\"o}m}  \&
  {Andersen}}{{Holmberg} et~al.}{2009}]{2009A&A...501..941H}
{Holmberg} J.,  {Nordstr{\"o}m} B.,   {Andersen} J.,  2009, \mn@doi [\aap]
  {10.1051/0004-6361/200811191}, \href
  {https://ui.adsabs.harvard.edu/abs/2009A&A...501..941H} {501, 941}

\bibitem[\protect\citeauthoryear{{Jao}, {Henry}, {Gies}  \& {Hambly}}{{Jao}
  et~al.}{2018}]{2018ApJ...861L..11J}
{Jao} W.-C.,  {Henry} T.~J.,  {Gies} D.~R.,   {Hambly} N.~C.,  2018, \mn@doi
  [\apjl] {10.3847/2041-8213/aacdf6}, \href
  {https://ui.adsabs.harvard.edu/abs/2018ApJ...861L..11J} {861, L11}

\bibitem[\protect\citeauthoryear{{Jarrett} et~al.,}{{Jarrett}
  et~al.}{2011}]{2011ApJ...735..112J}
{Jarrett} T.~H.,  et~al., 2011, \mn@doi [\apj] {10.1088/0004-637X/735/2/112},
  \href {https://ui.adsabs.harvard.edu/abs/2011ApJ...735..112J} {735, 112}

\bibitem[\protect\citeauthoryear{{Karovicova}, {White}, {Nordlander},
  {Casagrande}, {Ireland}  \& {Huber}}{{Karovicova}
  et~al.}{2022}]{2022A&A...658A..47K}
{Karovicova} I.,  {White} T.~R.,  {Nordlander} T.,  {Casagrande} L.,  {Ireland}
  M.,   {Huber} D.,  2022, \mn@doi [\aap] {10.1051/0004-6361/202141833}, \href
  {https://ui.adsabs.harvard.edu/abs/2022A&A...658A..47K} {658, A47}

\bibitem[\protect\citeauthoryear{{Kostov} et~al.,}{{Kostov}
  et~al.}{2020}]{2020AJ....159..253K}
{Kostov} V.~B.,  et~al., 2020, \mn@doi [\aj] {10.3847/1538-3881/ab8a48}, \href
  {https://ui.adsabs.harvard.edu/abs/2020AJ....159..253K} {159, 253}

\bibitem[\protect\citeauthoryear{{Lightkurve Collaboration}
  et~al.,}{{Lightkurve Collaboration} et~al.}{2018}]{2018ascl.soft12013L}
{Lightkurve Collaboration} et~al., 2018, {Lightkurve: Kepler and TESS time
  series analysis in Python}, Astrophysics Source Code Library (\mn@eprint
  {ascl} {1812.013})

\bibitem[\protect\citeauthoryear{{MacDonald} \& {Gizis}}{{MacDonald} \&
  {Gizis}}{2018}]{2018MNRAS.480.1711M}
{MacDonald} J.,  {Gizis} J.,  2018, \mn@doi [\mnras] {10.1093/mnras/sty1888},
  \href {https://ui.adsabs.harvard.edu/abs/2018MNRAS.480.1711M} {480, 1711}

\bibitem[\protect\citeauthoryear{{Mamajek} et~al.,}{{Mamajek}
  et~al.}{2015}]{2015arXiv151007674M}
{Mamajek} E.~E.,  et~al., 2015, arXiv e-prints, \href
  {https://ui.adsabs.harvard.edu/abs/2015arXiv151007674M} {p. arXiv:1510.07674}

\bibitem[\protect\citeauthoryear{{Martin} et~al.,}{{Martin}
  et~al.}{2019}]{2019A&A...624A..68M}
{Martin} D.~V.,  et~al., 2019, \mn@doi [\aap] {10.1051/0004-6361/201833669},
  \href {https://ui.adsabs.harvard.edu/abs/2019A&A...624A..68M} {624, A68}

\bibitem[\protect\citeauthoryear{{Maxted}}{{Maxted}}{2018}]{2018A&A...616A..39M}
{Maxted} P.~F.~L.,  2018, \mn@doi [\aap] {10.1051/0004-6361/201832944}, \href
  {https://ui.adsabs.harvard.edu/abs/2018A&A...616A..39M} {616, A39}

\bibitem[\protect\citeauthoryear{{Maxted}}{{Maxted}}{2023}]{2023MNRAS.522.2683M}
{Maxted} P. F.~L.,  2023, \mn@doi [\mnras] {10.1093/mnras/stad1112}, \href
  {https://ui.adsabs.harvard.edu/abs/2023MNRAS.522.2683M} {522, 2683}

\bibitem[\protect\citeauthoryear{{Maxted}, {Serenelli}  \&
  {Southworth}}{{Maxted} et~al.}{2015}]{2015A&A...575A..36M}
{Maxted} P.~F.~L.,  {Serenelli} A.~M.,   {Southworth} J.,  2015, \mn@doi [\aap]
  {10.1051/0004-6361/201425331}, \href
  {https://ui.adsabs.harvard.edu/abs/2015A&A...575A..36M} {575, A36}

\bibitem[\protect\citeauthoryear{{Maxted} et~al.,}{{Maxted}
  et~al.}{2022}]{2022MNRAS.513.6042M}
{Maxted} P.~F.~L.,  et~al., 2022, \mn@doi [\mnras] {10.1093/mnras/stac1270},
  \href {https://ui.adsabs.harvard.edu/abs/2022MNRAS.513.6042M} {513, 6042}

\bibitem[\protect\citeauthoryear{{Maxted}, {Triaud}  \& {Martin}}{{Maxted}
  et~al.}{2023}]{2023Univ....9..498M}
{Maxted} P. F.~L.,  {Triaud} A. H.~M.~J.,   {Martin} D.~V.,  2023, \mn@doi
  [Universe] {10.3390/universe9120498}, \href
  {https://ui.adsabs.harvard.edu/abs/2023Univ....9..498M} {9, 498}

\bibitem[\protect\citeauthoryear{{Miller}, {Maxted}  \& {Smalley}}{{Miller}
  et~al.}{2020}]{2020MNRAS.497.2899M}
{Miller} N.~J.,  {Maxted} P.~F.~L.,   {Smalley} B.,  2020, \mn@doi [\mnras]
  {10.1093/mnras/staa2167}, \href
  {https://ui.adsabs.harvard.edu/abs/2020MNRAS.497.2899M} {497, 2899}

\bibitem[\protect\citeauthoryear{{Munari} \& {Zwitter}}{{Munari} \&
  {Zwitter}}{1997}]{1997A&A...318..269M}
{Munari} U.,  {Zwitter} T.,  1997, \aap, \href
  {https://ui.adsabs.harvard.edu/abs/1997A&A...318..269M} {318, 269}

\bibitem[\protect\citeauthoryear{{Nelson} \& {Davis}}{{Nelson} \&
  {Davis}}{1972}]{1972ApJ...174..617N}
{Nelson} B.,  {Davis} W.~D.,  1972, \mn@doi [\apj] {10.1086/151524}, \href
  {https://ui.adsabs.harvard.edu/abs/1972ApJ...174..617N} {174, 617}

\bibitem[\protect\citeauthoryear{{Onken}, {Wolf}, {Bessell}, {Chang}, {Luvaul},
  {Tonry}, {White}  \& {Da Costa}}{{Onken} et~al.}{2024}]{2024arXiv240202015O}
{Onken} C.~A.,  {Wolf} C.,  {Bessell} M.~S.,  {Chang} S.-W.,  {Luvaul} L.~C.,
  {Tonry} J.~L.,  {White} M.~C.,   {Da Costa} G.~S.,  2024, \mn@doi [arXiv
  e-prints] {10.48550/arXiv.2402.02015}, \href
  {https://ui.adsabs.harvard.edu/abs/2024arXiv240202015O} {p. arXiv:2402.02015}

\bibitem[\protect\citeauthoryear{{Pepe} et~al.,}{{Pepe}
  et~al.}{2002}]{2002Msngr.110....9P}
{Pepe} F.,  et~al., 2002, The Messenger, \href
  {https://ui.adsabs.harvard.edu/abs/2002Msngr.110....9P} {110, 9}

\bibitem[\protect\citeauthoryear{{Pepe} et~al.,}{{Pepe}
  et~al.}{2021}]{2021A&A...645A..96P}
{Pepe} F.,  et~al., 2021, \mn@doi [\aap] {10.1051/0004-6361/202038306}, \href
  {https://ui.adsabs.harvard.edu/abs/2021A&A...645A..96P} {645, A96}

\bibitem[\protect\citeauthoryear{{Pollacco} et~al.,}{{Pollacco}
  et~al.}{2006}]{2006PASP..118.1407P}
{Pollacco} D.~L.,  et~al., 2006, \mn@doi [\pasp] {10.1086/508556}, \href
  {https://ui.adsabs.harvard.edu/abs/2006PASP..118.1407P} {118, 1407}

\bibitem[\protect\citeauthoryear{{Queiroz} et~al.,}{{Queiroz}
  et~al.}{2018}]{2018MNRAS.476.2556Q}
{Queiroz} A.~B.~A.,  et~al., 2018, \mn@doi [\mnras] {10.1093/mnras/sty330},
  \href {https://ui.adsabs.harvard.edu/abs/2018MNRAS.476.2556Q} {476, 2556}

\bibitem[\protect\citeauthoryear{{Rabus} et~al.,}{{Rabus}
  et~al.}{2019}]{2019MNRAS.484.2674R}
{Rabus} M.,  et~al., 2019, \mn@doi [\mnras] {10.1093/mnras/sty3430}, \href
  {https://ui.adsabs.harvard.edu/abs/2019MNRAS.484.2674R} {484, 2674}

\bibitem[\protect\citeauthoryear{{Ricker} et~al.,}{{Ricker}
  et~al.}{2015}]{2015JATIS...1a4003R}
{Ricker} G.~R.,  et~al., 2015, \mn@doi [Journal of Astronomical Telescopes,
  Instruments, and Systems] {10.1117/1.JATIS.1.1.014003}, \href
  {https://ui.adsabs.harvard.edu/abs/2015JATIS...1a4003R} {1, 014003}

\bibitem[\protect\citeauthoryear{{Sebastian} et~al.,}{{Sebastian}
  et~al.}{2024}]{2024MNRAS.tmp.1016S}
{Sebastian} D.,  et~al., 2024, \mn@doi [\mnras] {10.1093/mnras/stae459}, \href
  {https://ui.adsabs.harvard.edu/abs/2024MNRAS.tmp.1016S} {}

\bibitem[\protect\citeauthoryear{{Serenelli}, {Bergemann}, {Ruchti}  \&
  {Casagrande}}{{Serenelli} et~al.}{2013}]{2013MNRAS.429.3645S}
{Serenelli} A.~M.,  {Bergemann} M.,  {Ruchti} G.,   {Casagrande} L.,  2013,
  \mn@doi [\mnras] {10.1093/mnras/sts648}, \href
  {http://adsabs.harvard.edu/abs/2013MNRAS.429.3645S} {429, 3645}

\bibitem[\protect\citeauthoryear{{Skrutskie} et~al.,}{{Skrutskie}
  et~al.}{2006}]{2006AJ....131.1163S}
{Skrutskie} M.~F.,  et~al., 2006, \mn@doi [\aj] {10.1086/498708}, \href
  {https://ui.adsabs.harvard.edu/abs/2006AJ....131.1163S} {131, 1163}

\bibitem[\protect\citeauthoryear{{Soubiran} et~al.,}{{Soubiran}
  et~al.}{2024}]{2024A&A...682A.145S}
{Soubiran} C.,  et~al., 2024, \mn@doi [\aap] {10.1051/0004-6361/202347136},
  \href {https://ui.adsabs.harvard.edu/abs/2024A&A...682A.145S} {682, A145}

\bibitem[\protect\citeauthoryear{{Southworth}}{{Southworth}}{2010}]{2010MNRAS.408.1689S}
{Southworth} J.,  2010, \mn@doi [\mnras] {10.1111/j.1365-2966.2010.17231.x},
  \href {https://ui.adsabs.harvard.edu/abs/2010MNRAS.408.1689S} {408, 1689}

\bibitem[\protect\citeauthoryear{{Southworth}}{{Southworth}}{2015}]{2015ASPC..496..164S}
{Southworth} J.,  2015, in {Rucinski} S.~M.,  {Torres} G.,   {Zejda} M.,  eds,
  Astronomical Society of the Pacific Conference Series Vol. 496, Living
  Together: Planets, Host Stars and Binaries. p.~164 (\mn@eprint {arXiv}
  {1411.1219})

\bibitem[\protect\citeauthoryear{{Southworth}}{{Southworth}}{2023}]{2023Obs...143...71S}
{Southworth} J.,  2023, \mn@doi [The Observatory] {10.48550/arXiv.2301.02531},
  \href {https://ui.adsabs.harvard.edu/abs/2023Obs...143...71S} {143, 71}

\bibitem[\protect\citeauthoryear{{Standing} et~al.,}{{Standing}
  et~al.}{2023}]{2023NatAs...7..702S}
{Standing} M.~R.,  et~al., 2023, \mn@doi [Nature Astronomy]
  {10.1038/s41550-023-01948-4}, \href
  {https://ui.adsabs.harvard.edu/abs/2023NatAs...7..702S} {7, 702}

\bibitem[\protect\citeauthoryear{{Triaud} et~al.,}{{Triaud}
  et~al.}{2017}]{2017A&A...608A.129T}
{Triaud} A. H.~M.~J.,  et~al., 2017, \mn@doi [\aap]
  {10.1051/0004-6361/201730993}, \href
  {https://ui.adsabs.harvard.edu/abs/2017A&A...608A.129T} {608, A129}

\bibitem[\protect\citeauthoryear{{Weiss} \& {Schlattl}}{{Weiss} \&
  {Schlattl}}{2008}]{2008Ap&SS.316...99W}
{Weiss} A.,  {Schlattl} H.,  2008, \mn@doi [\apss] {10.1007/s10509-007-9606-5},
  \href {https://ui.adsabs.harvard.edu/abs/2008Ap&SS.316...99W} {316, 99}

\makeatother
\end{thebibliography}


\bsp	
\label{lastpage}
\end{document}